\documentclass[pre,twocolumn,superscriptaddress,amsmath,amssymb,floatfix]{revtex4}

\usepackage{graphicx}
\usepackage{bm}
\begin{document}

\title{Coarse-grained interaction potentials for polyaromatic hydrocarbons}

\author{O. Anatole von Lilienfeld}
\email{vonlilienfeld@gmail.com}
\homepage{http://lcbcpc21.epfl.ch/Group_members/anatole/}
\affiliation{Department of Chemistry, New York University, New York, NY 10003, USA}
\affiliation{Institute for Pure and Applied Mathematics,
 460 Portola Plaza, University of California Los Angeles, Los Angeles,
 CA 90095-7121, USA}

\author{Denis Andrienko}
\email{denis.andrienko@mpip-mainz.mpg.de}
\homepage{http://www.mpip-mainz.mpg.de/~andrienk/}
\affiliation{Institute for Pure and Applied Mathematics,
 460 Portola Plaza, University of California Los Angeles, Los Angeles,
 CA 90095-7121, USA}
\affiliation{Max-Planck-Institut f\"{u}r Polymerforschung, Ackermannweg 10, 55128 Mainz, Germany}

\date{\today}

\begin{abstract}
 Using Kohn-Sham density functional theory (KS-DFT), we have studied the interaction between
 various polyaromatic hydrocarbon molecules. The systems range from
 mono-cyclic benzene up to hexabenzocoronene (hbc). For several conventional
 exchange-correlation functionals potential energy curves of interaction of the
 $\pi$-$\pi$ stacking hbc dimer are reported. It is found that
 all pure local density or generalized gradient approximated functionals
 yield qualitatively incorrect predictions regarding structure and
 interaction. Inclusion of a non-local, atom-centered
 correction to the KS-Hamiltonian enables quantitative predictions.
 The computed potential energy surfaces of
 interaction yield parameters for a coarse-grained potential, which can be employed to study
 discotic liquid-crystalline mesophases of derived polyaromatic macromolecules.
\end{abstract}

\maketitle

\section{Introduction}
Discotic thermotropic liquid crystals can be formed by flat molecules
with a central aromatic core and several aliphatic chains attached at
the edges~\cite{ChandrasekharR90,BushbyL02}. The size and the shape of
the cores can be varied, as well as the length and the structure of the
side chains, which allows the control of functional properties of these
mesophases~\cite{BrandKIM00,Kumar04,MullerKM98}.
Liquid-crystalline properties such as fluidity are of help to process
these compounds and even to develop self-healing materials.

The self-organization of the aromatic cores into $\pi$-$\pi$ electron
bonded stacks surrounded by saturated hydrocarbons allows
one-dimensional charge transport along the
columns~\cite{vandeCraatsWFBHM99,Schmidt-MendeFMMFM01}.
Unfortunately, the spatial arrangement of stacks is not perfect,
i.e.~the columns can be misaligned, tilted, or form various types of
topological defects.  In addition, the local alignment of molecules in
columns can vary for different compounds.  This
considerably affects the intermolecular overlap of the $\pi$
orbitals and thereby the efficiency of the charge transport in a
single column.
%
As a consequence, the details of the morphology of the
conducting film are crucial~\cite{HallsAMWIWF00,ShaheenBSPFH01,AriasMSHIWRF01}.
An accurate \textit{in silico} prediction of mesophase properties prior to the actual
synthesis of the compounds could account for a considerable gain in
efficiency on the route towards the design of macromolecular photo
devices~\cite{DebijePdWTSWM04}.

An accurate understanding of the constituting molecules and
their intermolecular interactions is mandatory
for controling the local alignment of the disks or
the global arrangement of the columns in the mesophase.
%
Depending on the chosen length- and time-scales different
methods can be considered. In principle, at the quantum chemistry
level, one can study electronic, inter and intra molecular adsorption and adhesion
processes~\cite{AIMD_PNAS_TUCKERMAN2005}, and even compound
design~\cite{myself-prl2005} without empiricism.
Still at an atomistic resolution - but using
empirical molecular force-fields - nanometer and nanosecond simulations
can yield local properties, such as order parameters, or molecular
arrangements~\cite{Maliniak92,OnoK92,MulderSPKdSK03,CinacchiCT04}.
In an even more extended time and length scale limit ($\mu$m and $\mu$s),
coarse-grained simulations allow to describe the morphology of bulk material,
global arrangement of macromolecular objects such as columns or generic
phase diagrams, and
defects~\cite{VeermanF92,EmersonLW94,BatesL96,Zewdie98,CinacchiT02,CaprionBR03,Bellier-CastellaCR04}.

To our knowledge, discotic materials have been studied only very little and if so
with idealized model potentials. The reason being that an {\em ab initio}
treatment of the dispersion forces is computationally very demanding.
Molecular dynamics (MD) is able to treat larger systems, however the details of the electronic structure, which are crucial for
an understanding of electron transport, are by construction not included
in MD. Moreover, it requires empirical parameters for the atom-atom interactions, and
 defects and the mesophase morphology can only be studied at even more coarse levels.
Consequently, multiscale methodologies~\cite{Baschnagel00,AbramsDK03,DelleSiteAAK02,AbramsK03,zhou2005a,zhou:2005.b,andrienko:2005.b} seem to represent the most adequate
and tractable description of these
systems.

The aim of this work is to make first steps towards multiscale
modeling of discotic mesophases of polyaromatic hydrocarbons.
Namely, coarse-grained potentials for the interaction
between representative polyaromatic molecules in a face-to-face geometry
are obtained from first principles.
They can be applied to the study of macroscopic properties
of these materials or their chemically derived structures.

\section{Computational Details}
The intermolecular attraction between polyaromatic systems
is partly attributed to London-dispersion forces. These forces
result from the correlated fluctuation of non-overlapping
electron densities of molecular fragments~\cite{MQED_CRAIG}.
Their prediction from first principles has remained a long-standing challenge
because of the very high accuracy required to describe
electron correlation effects. Explicitly correlated wavefunction
methods such as coupled-cluster, configuration interaction, or
quantum Monte Carlo allow for an accurate treatment of
these forces but are computationally prohibitively expensive for all
but the smallest polyaromatic hydrocarbons, such as for instance
the benzene dimer~\cite{TsuzukisBenzeneDimer}.
Kohn-Sham density functional theory (KS-DFT), on the other hand,
would be an exact electronic structure method if the
true exchange-correlation (xc) term in the KS-potential was
known. Unfortunately, this is not the case and for all practical purposes
approximations have to be made. While some of the pure xc functionals
fortuitously but inconsistently predict binding for London-dispersion
complexes, it is not yet generally possible to describe correctly vdW
interactions within DFT using the local density approximation (LDA),
the generalized gradient approximation (GGA) or even the - on average
more accurate - hybrid exchange-correlation
functionals~\cite{ChemistsGuidetoDFT,Pulay-NoVdw,Becke-NoVdw,Sprik-NoVdw,CriticalNote_DFT4He,ScolesVDW}.

Considering the ubiquitous nature of these intermolecular forces and
their importance for self-assembly, much effort is being devoted to
design superior xc-potentials which can account correctly for all
intermolecular interactions. The use of nonlocal correlations
by electron density partitioning~\cite{Wesolowski-VDW1,Wesolowski-VDW2}
can efficiently remedy this deficiency but implies an \textit{a priori}
assignment of molecular fragments.
A `van der Waals' functional as proposed in
Ref.~\cite{nonlocalVDW1,nonlocalVDW2,nonlocalVDW3,nonlocalVDW4,nonlocalVDW5},
and based on response theory becomes rapidly intractable, such as
the schemes described in Ref.~\cite{SzalewiczVDW,KohnVdwDFT}.
As a consequence, empirical \textit{a posteriori} pairwise atom-atom
based correction terms~\cite{Sprik-NoVdw, LeSar-VDW} to the energy are in
wide spread use for practical applications. The required parameters
and damping functions for the correct repulsive behavior can be
obtained from experiments or from various theoretical
approaches including time dependent DFT~\cite{Grimme2004, BaerendsC6,
BeckesC6,BechstedtPRL2005}. However, these $r^{-6}$-dependent
corrections to the energy and ionic forces need artificial damping
functions to allow for the correct repulsive behavior, and more
importantly leave the electronic structure uncorrected.

In this study, London-dispersion forces are computed
from an improved electronic structure calculation which
exploits a recently presented
semi-empirical dispersion calibrated atom-centered (DCACP)
correction, $\hat{v}^{disp}_i$, to a given Hamiltonian~\cite{myself-prl2004}.
Specifically, it can be seen as a nonlocal extension
of a given, local for LDA or GGA, xc-potential,
\begin{eqnarray}
\hat{v}^{extended}_{xc}(\mathbf{r}) &  = & \hat{v}_{xc}(\mathbf{r}) +
\sum_i \hat{v}_i^{disp}(\mathbf{r,r}',\{\lambda\}),
\end{eqnarray}
where index $i$ enumerates all atoms.
As generally suggested in Ref.~\cite{myself-jcp2004}, the
atom-type-dependent parameters $\{\lambda\}$ require preliminary calibration
for an improved electronic structure fulfilling additional
requirements, such as exerting a London-dispersion force on
the ions. Specifically, the BLYP-DCACP for carbon as it has
been introduced, calibrated to the benzene dimer, and assessed in
Refs.~\cite{myself-prl2004,myself-prb2005,myself-enrico2005} is employed.
Generalization of this correction to other xc-functionals than BLYP has
already been carried out~\cite{myself-mauricio2006}.


\begin{figure}
\includegraphics[width=8cm]{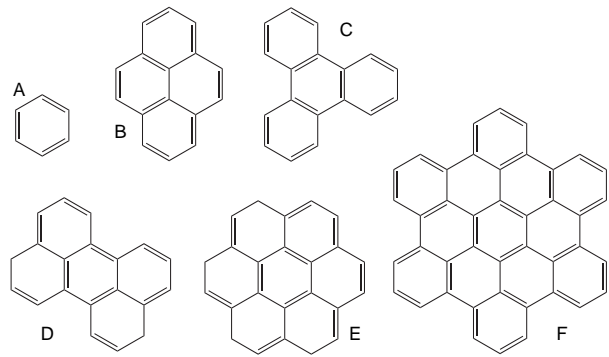}
\caption{
Sketches of the polycyclic aromatic hydrocarbons benzene (A), pyrene
(B), triphenylene (C), perylene (D), coronene (E), and
hexabenzocoronene (hbc) (F). All hydrogens are omitted for the sake of
clarity.  Bonds are represented by edges, carbon atoms by vertices.
}\label{fig:structures}
\end{figure}
All DFT calculations have been carried out using the plane-wave
basis set electronic structure program
\textsc{cpmd 3.92}~\cite{cpmd3.92}, the xc-functionals
BLYP~\cite{B88X,ColleSalvetti,lyp}, BP~\cite{B88X,perdew-BP}, PBE~\cite{PBE},
LDA (using the Perdew and Zunger fit~\cite{perdew-xc} to the
data of Ceperley and Alder~\cite{ceperley-xc}),
Goedecker pseudopotentials from Refs.~\cite{SG,sgpsp,krackPP},
and a plane-wave cutoff of 100 Ry. The isolated
system module in \textsc{cpmd} has been employed together
with the Poisson-solver of Tuckerman and
Martyna~\cite{martyna-tuckerman}. The box-size is
sufficiently converged at 19$\times$19$\times$20
{\AA}$^3$ for the largest system (hbc) and has been
kept fixed for all molecules and all distances.
Carbon-type DCACPs~\cite{myself-prl2004} have been used only as a correction to the
BLYP functional, no correction has been employed for hydrogen atoms.
For the calibration of the DCACP's in Ref.~\cite{myself-prl2004},
the M{\o}ller-Plesset second order perturbation theory
energy of interaction of the benzene dimer was used as a reference.
Since the plane wave basis is independent of atomic positions,
no basis set superposition errors occurr. Relative
energies have been computed for identical box-sizes and
cutoffs. For all geometry optimizations the residual
tolerance for ionic forces has been set to
0.0005 a.u. For all calculations of energies of interaction, the monomer
geometries have been optimized with the given xc-functional.
Thereafter, for the calculation of the interaction curves, the
intramolecular geometries of the top-on-top moieties
have been hold fixed and only the intermolecular distance
has been varied.

Several aromatic disks with symmetric cores have been
selected. Sketches of their chemical composition
are shown in Fig.~\ref{fig:structures}. Namely, complexes
of benzene (A), pyrene (B), triphenylene (C), perylene (D), coronene (E), and
hexabenzocoronene (hbc) (F) have been studied, which all fulfill
H\"uckel's $(4N+2)$ rule.

\begin{figure}
\includegraphics[width=8cm]{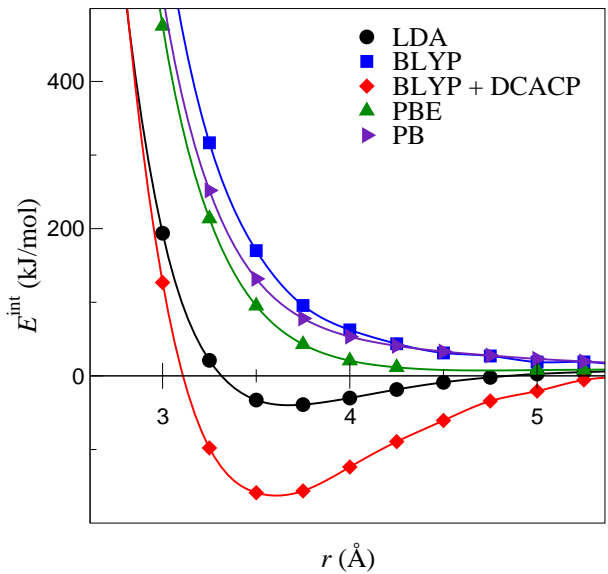}
\caption{\label{FIG2a}
Total potential energy of interaction
for hexabenzocoronene (see structure F, Fig.~\ref{fig:structures}) as a function of intermolecular distance for
the LDA, BLYP, PBE, BP, and BLYP + DCACP xc-functionals.
}\end{figure}

\section{Results and discussion}
\subsection{Different density functionals}
First, several xc functionals have been assessed by evaluating
the potential energies of interaction,
\begin{equation}
E^{\rm int} = E^{\rm dimer} - 2E^{\rm monomer},
\end{equation}
for hexabenzocoronene using LDA, BLYP, PBE, and BP. The results are shown in Fig.~\ref{FIG2a}.
As expected, LDA exhibits some fortuitous binding for graphite-like structures,
however when using the more sophisticated GGA-functionals, a completely repulsive behavior is obtained.
It is found that the PBE gives the least repulsive interaction, followed by BP, and BLYP,
suggesting that Becke's exchange potential is too repulsive for the investigated systems.
Upon inclusion of the DCACP-extension into the BLYP-DFT Hamiltonian,
the interaction energy curve is in a reasonable agreement with what
can be expected from experimental results, available for coronene~\cite{ZachariaUH04}.

\subsection{Interaction energy profiles}
\begin{figure}
\includegraphics[width=8cm]{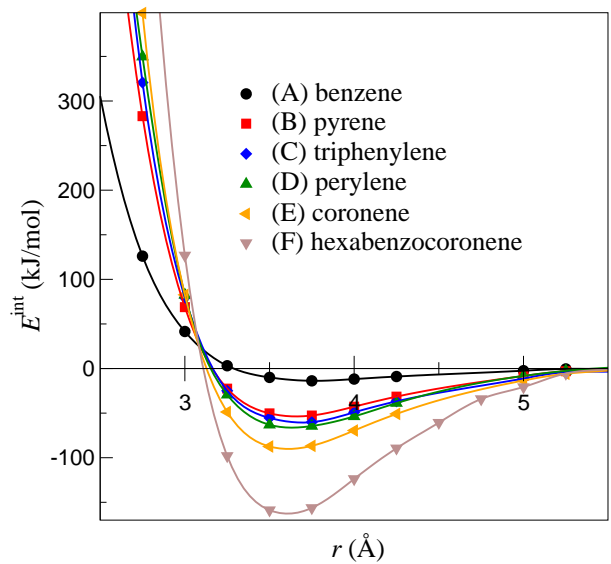}
\caption{\label{FIG2b}
Energies of interaction vs.~molecule-molecule separation for all studied
aromatic systems. Calculations are performed using BLYP+DCACP functional.}
\end{figure}

The interaction energy profiles $E^{int}(r)$ have been calculated for all systems
in the face-to-face geometry using the BLYP-DCACP KS-Hamiltonian.
The results are presented in Fig.~\ref{FIG2b}.
The calculated profiles are interpolated with cubic splines. The equilibrium
separations, $r_{eq}$ and the minima of the interaction energy, $E^{int}_{eq}$,
are determined from the interpolated curves and are reported in
Table~\ref{TAB1}, together with the value for two isolated
graphene sheets from Ref.~\cite{myself-prl2004}.

When increasing the disk-size of the systems, the energy of
interaction per carbon atom increases; correspondingly, the equilibrium
separation decreases. The remaining difference of the largest system with respect to graphene
is most probably due to the static multipole of the saturating hydrogen atoms,
which is known to represent up to 7\% of the interaction energy in the
case of benzene~\cite{Contreras-CamachoUBM04}.
The convergence of the interaction parameters with the system's size
can be exploited for extrapolations to even larger structures,
such as supernaphtalene or supertriphenylene~\cite{WatsonFM01}, for which
the importance of symmetry and hydrogen atoms can be expected to decrease, due to the
even smaller ratio between molecular perimeter and surface.

\begin{table}[h]
\caption{
Minimum of the interaction energy, $E^{int}_{eq}$
together with the equilibrium distance, $r_{eq}$, for all the systems. \textit{graph} corresponds to
the calculated prediction for two graphene sheets in vacuum
which compares well to experiment.}
\begin{ruledtabular}
\begin{tabular}{lccc}
System & $N$ & -$E^{int}_{eq} $  (kJ/mol) & $r_{eq}$ ({\AA})\\\hline
(A) benzene &6 & 13.8  & 3.77 \\
(B) pyrene &16& 53.7  & 3.67 \\
(C) triphenylene &18& 60.6  & 3.70 \\
(D) perylene &20& 66.3  & 3.63 \\
(E) coronene &24& 90.1  & 3.60 \\
(F) hexabenzocoronene &42& 162.7  & 3.60 \\
graph\footnotemark[1]&42& 140.7 & 3.30 \\
\end{tabular}
\end{ruledtabular}
\footnotetext[1] {from Ref.~\cite{myself-prl2004}}
\label{TAB1}
\end{table}

\begin{figure}
\includegraphics[width=8cm]{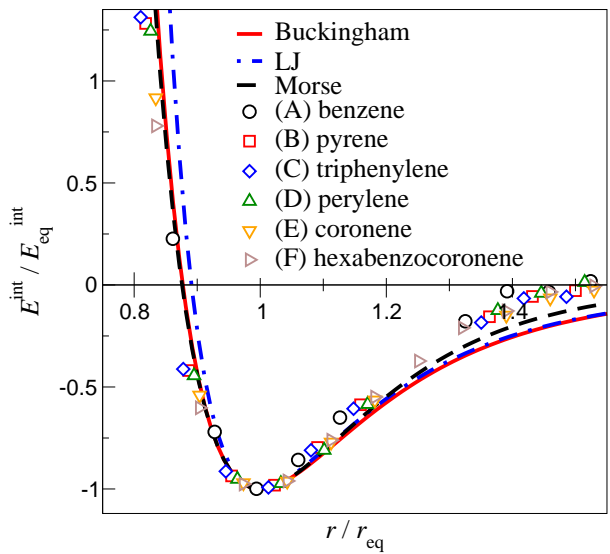}
\caption{
Interaction energies in units of $E^{int}_{eq}$ vs.~the separation scaled by $r_{eq}$. The
DCACP-BLYP-DFT results (symbols) for all systems fall on a single master curve.
Solid lines represent fits corresponding to different
coarse-grained potentials, Lennard-Jones, Morse, and Buckingham.}
\label{fig:scaling}
\end{figure}

To obtain a coarse-grained interaction potential we have normalized the energies by the interaction energy at
the equilibrium distance, $E^{int}_{eq}$, and have scaled the molecule-molecule separation with the equilibrium
distance, $r_{eq}$. After scaling all DFT profiles superimpose on a single curve,
as illustrated in Fig.~\ref{fig:scaling}. This suggests that the main
contribution to the non-bonded interactions of the atoms in the dimer
is due to an (additive) pairwise carbon-carbon interaction.

We have fitted the master curve with three frequently used
potentials, Lennard-Jones (LJ), Morse, and Buckingham.
All the potentials have been constrained to have the minimum of the energy,
$u_{eq} = -1$, at the dimensionless separation, $x_{eq} = 1$.
Under this constraint, the Lennard-Jones potential
is parameter-free, while the Morse and Buckingham potentials
have each a single fitting parameter,
\begin{eqnarray}
u_{\rm LJ}(x) &=& \left[ \frac{1}{x^{12}} - \frac{2}{x^6} \right], \nonumber\\
u_{\rm Morse}(x) &=& \left[ 1-e^{-\alpha(x-1)} \right]^2 - 1, \nonumber\\
u_{\rm Buckingham} &=& \frac{1}{\beta - 6} \left[ 6 e^{-\beta(x-1)} -
\frac{\beta}{x^6}\right].
\label{eq:potentials}
\end{eqnarray}
Here, $x = r / r_{eq}$. The interval $x = [0.9,1.4]$ has been used to
determine the fitting parameters $\alpha = 5.6$  and $\beta = 12.3$.

The dimensional potential is obtained by
multiplying $u(x)$ with the absolute value of $E^{int}_{eq}$ and using
$x = r / r_{eq}$ as the argument. The corresponding values are given in
Table~\ref{TAB1}.
For instance, the Lennard-Jones potential for benzene would have a
potential energy of interaction of $13.8 \times u_{\rm LJ}(r/3.77)$,
in kJ/mol, and $r$ in {\AA}.

All the three potentials fit very well to the DFT calculations. 
The Morse potential reproduces best the repulsive and the attractive part of the potential.
The fact that all interaction potentials fall on a single
master-curve implies that the performance of the fits remains
constant for all investigated supermolecular systems.

\subsection{Coarse-grained potentials}
\subsubsection{United atoms model}
Equations~(\ref{eq:potentials}), can, in principle,
be used directly for the parametrization of interaction
potentials treating the whole molecule as one
interacting point, such as the Gay-Berne potential.
Here, however, we will be interested in more accurate representations.
We start with a united atom model, in which all hydrogen atoms are
embedded into the carbons they saturate.
All considered molecules are assumed to
be rigid, i.e., no stretching, bending, or torsional energy is included.
Since, already for benzene, the electrostatic contribution
represents up to 7\% of the total intermolecular energy~\cite{Contreras-CamachoUBM04},
we have neglected this contribution for the parametrization of the coarse-grained model.

For atomistic two-body potentials the interaction of two
molecules is a sum of the corresponding pair interactions of all atoms.
If the effective dispersion-repulsion interaction between two
atoms (or united atoms) $i$ and $j$ of different molecules is
represented by a Lennard-Jones 6-12 potential,
\begin{equation}
U^{ij}({r_{ij}}) = 4 \epsilon_{ij} \left[ \left(  \frac{\sigma_{ij}} {{r_{ij}}} \right)^{12} - \left(\frac{\sigma_{ij}} {{r_{ij}}} \right)^{6} \right],
\end{equation}
then the molecule-molecule interaction is the sum over all pair interactions
of the atoms belonging to different molecules,
\begin{equation}
U(1,2) = \sum_{i \in 1 ,j \in 2} U^{ij}({r_{ij}}).
\label{eq:mol}
\end{equation}

In what follows we have assumed that all \textit{inner}
and \textit{edge} carbons have the same parameters, $\epsilon$ and $\sigma$,
and optimized these parameters to reproduce the desired molecule-molecule interaction.
For the fit of the DFT data to Eq.~\ref{eq:mol}, we have considered only the
region close to the equilibrium separation $r_{eq}$.
This limitation is due to the fact that
the employed DCACP-correction to the DFT functional~\cite{myself-prl2004}
was calibrated to reproduce only this equilibrium region, and does
not explicitly include the typical dissociative $r^{-6}$-behavior.

\begin{figure}
\includegraphics[width=8cm]{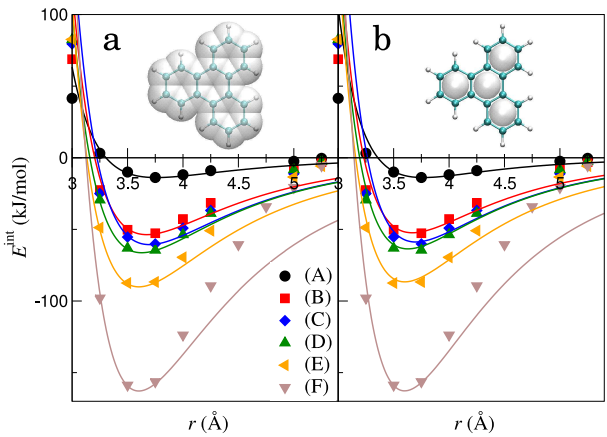}
\caption{DFT-data and corresponding fitting curves for united atoms (a) and benzene-bead (b)
representations. The insets illustrate the representations for triphenylene (C).}
\label{fig:fit_LJ}
\end{figure}

The resulting parameters of the fit are summarized in Tab.~\ref{TAB2}. The
DFT data points, together with the corresponding fitting curves are shown in Fig~\ref{fig:fit_LJ}(a).
Again, the Lennard-Jones potential does not reproduce perfectly well
the attractive tail but, as explained before, the position of the minimum is more important
for our purposes.

\begin{table}
\begin{ruledtabular}
\begin{tabular}{l|ccc|ccc}
System & $N$   & $\epsilon$       & $\sigma$
       & $N_B$ & $ \epsilon_B $   & $\sigma_B$  \\\hline
(A) benzene &6 & 0.466  &  3.556 & 1 & 13.802 & 3.358 \\
(B) pyrene &16& 0.408   &  3.541 & 4 & 5.019  & 3.428 \\
(C) triphenylene &18& 0.407  &  3.569 & 4 & 6.963 & 3.417\\
(D) perylene &20& 0.383  &  3.512 & 4 & 4.984 & 3.390 \\
(E) coronene &24& 0.393   &  3.498 & 7 & 3.536 & 3.413 \\
(F) hbc &42& 0.353 & 3.518 & 13 & 2.873 & 3.459\\
\end{tabular}
\end{ruledtabular}
\caption{Lennard-Jones parameters for united atom ($\epsilon$,$\sigma$) and benzene-bead
($\epsilon_B$,$\sigma_B$) parameterizations of all the studied systems.
The cutoff $r_{cut} = 15\,$ {\AA} has been used to evaluate the
potential in Eq.~(\ref{eq:mol}). Energies $\epsilon$, $\epsilon_{B}$ are
in kJ/mol; $\sigma$ and $\sigma_B$ are in {\AA}. $N$ is the number of
carbons, $N_B$ is the number of benzene beads per molecule.}
\label{TAB2}
\end{table}

We can also compare some of the obtained parameters to values of existing force fields.
For instance, a number of united atom force fields are available for
benzene~\cite{WickMS00,ErringtonP99,Linse84,LinseKJ84,ClaessensFR83,EvansW76}.
Depending on the employed parametrization, the literature values are
$\sigma \in [3.25-3.57]$ {\AA} and $\epsilon \in [0.4-0.75]\, {\rm kJ/mol}$,
i.~e. the values of this study fall into the range of parameters predicted
from thermodynamic properties of benzene.

For the investigated systems, no abundant experimental data, particularly concerning
dimer interactions, is available. However, adsorption energies
of benzene or coronene on basal planes of graphite were measured~\cite{ZachariaUH04},
$\approx$ 50 kJ/mol and $\approx$ 120 kJ/mol, respectively.
While we find no quantitative agreement with the results for benzene,
the agreement for coronene is better. It is plausible that due to the interaction of
a molecule with the bulk of graphite, the adsorption energy on graphite is larger
than the interaction energy between two isolated benzene molecules.

\subsubsection{Benzene-bead representation}
In coarse-grained simulations, one frequently encounters the approximation that
fragments building up larger systems are rigid, i.e. their internal stretching, bending,
or torsional energy contributions are neglected.
Exploiting this assumption, a computationally more efficient coarse-grained
representation can be proposed in which each benzene is
represented as an interacting point located at its center of mass.
This reduces the number of degrees of freedom considerably. Assuming, likewise, a
Lennard-Jones type of interaction, the above presented procedure to fit to the DFT-data
has been applied to obtain the corresponding parameters
for benzene-beads. The results and fitting curves are displayed in
TABLE~\ref{TAB2} and Fig.~\ref{fig:fit_LJ}(b), respectively.
There is no significant difference in the interaction profiles
upon use of the united atom or the benzene-bead representation.
However, for the latter the size of the beads is relatively small, i.e.
only as large as the internal diameter of a benzene unit. This
implies that only mesophases in which molecules experience
interactions via a face-to-face arrangement can be studied.
Another limitation is that the rotational profile of the interaction energy,
e.g.~due to azimuthal rotation of a moiety, has not been included in the parameterizations.
Consequently, an accurate prediction of the helix structure, often observed
in columnar mesophases, can not be expected.

\section{Conclusions}
The DFT KS-potentials using LDA, BP, PBE, or BLYP approximations
to the xc-potential fail to correctly predict an attractive interaction
between polyaromatic hydrocarbon molecules.
The BLYP-DCACP-extension of Ref.~\cite{myself-prl2004}
has successfully been applied to compute interaction energies
for hexabenzocoronene, coronene, perylene,
triphenylene, pyrene, and benzene,
without any computational overhead or necessity of
{\em a priori} assignments of fragments.
By scaling the obtained data with equilibrium energy and
distance values, a single function has been found
to describe all molecule-molecule interactions
independent of the number of atoms.

The DFT results have been used to parameterize
a united atom representation, taking each carbon as an interacting site.
Additionally, another coarse-grained representation has
been proposed and parameterized in which each benzene unit represents
one Lennard-Jones bead.

The obtained interaction potentials will be
of use for future studies of columnar phases of
corresponding compounds or their derivatives within
atomistic molecular dynamics simulations or more coarse
representations.

\begin{acknowledgments}
Both authors would like to thank the organizers of
the workshop {\em Bridging Time and Length Scales in Materials Science
and Bio-Physics} which was held at the Institute for Pure and Applied
Mathematics at the University of California Los Angeles and which
inspired this study. Furthermore, the authors thank L.~Delle Site 
and K.~Kremer for helpful discussions and support.
OAvL is grateful for support from the SNF, Grant No.~PBEL2-110243.
\end{acknowledgments}
\bibliography{./article}

\begin{thebibliography}{79}
\expandafter\ifx\csname natexlab\endcsname\relax\def\natexlab#1{#1}\fi
\expandafter\ifx\csname bibnamefont\endcsname\relax
  \def\bibnamefont#1{#1}\fi
\expandafter\ifx\csname bibfnamefont\endcsname\relax
  \def\bibfnamefont#1{#1}\fi
\expandafter\ifx\csname citenamefont\endcsname\relax
  \def\citenamefont#1{#1}\fi
\expandafter\ifx\csname url\endcsname\relax
  \def\url#1{\texttt{#1}}\fi
\expandafter\ifx\csname urlprefix\endcsname\relax\def\urlprefix{URL }\fi
\providecommand{\bibinfo}[2]{#2}
\providecommand{\eprint}[2][]{\url{#2}}

\bibitem[{\citenamefont{Chandrasekhar and Ranganath}(1990)}]{ChandrasekharR90}
\bibinfo{author}{\bibfnamefont{S.}~\bibnamefont{Chandrasekhar}}
  \bibnamefont{and} \bibinfo{author}{\bibfnamefont{G.~S.}
  \bibnamefont{Ranganath}}, \bibinfo{journal}{Reports On Progress In Physics}
  \textbf{\bibinfo{volume}{53}}, \bibinfo{pages}{57} (\bibinfo{year}{1990}).

\bibitem[{\citenamefont{Bushby and Lozman}(2002)}]{BushbyL02}
\bibinfo{author}{\bibfnamefont{R.~J.} \bibnamefont{Bushby}} \bibnamefont{and}
  \bibinfo{author}{\bibfnamefont{O.~R.} \bibnamefont{Lozman}},
  \bibinfo{journal}{Current Opinion In Colloid and Interface Science}
  \textbf{\bibinfo{volume}{7}}, \bibinfo{pages}{343} (\bibinfo{year}{2002}).

\bibitem[{\citenamefont{Brand et~al.}(2000)\citenamefont{Brand, Kubel, Ito, and
  Mullen}}]{BrandKIM00}
\bibinfo{author}{\bibfnamefont{J.~D.} \bibnamefont{Brand}},
  \bibinfo{author}{\bibfnamefont{C.}~\bibnamefont{Kubel}},
  \bibinfo{author}{\bibfnamefont{S.}~\bibnamefont{Ito}}, \bibnamefont{and}
  \bibinfo{author}{\bibfnamefont{K.}~\bibnamefont{Mullen}},
  \bibinfo{journal}{Chemistry of Materials} \textbf{\bibinfo{volume}{12}},
  \bibinfo{pages}{1638} (\bibinfo{year}{2000}).

\bibitem[{\citenamefont{Kumar}(2004)}]{Kumar04}
\bibinfo{author}{\bibfnamefont{S.}~\bibnamefont{Kumar}},
  \bibinfo{journal}{Liquid Crystals} \textbf{\bibinfo{volume}{31}},
  \bibinfo{pages}{1037} (\bibinfo{year}{2004}).

\bibitem[{\citenamefont{Muller et~al.}(1998)\citenamefont{Muller, Kubel, and
  Mullen}}]{MullerKM98}
\bibinfo{author}{\bibfnamefont{M.}~\bibnamefont{Muller}},
  \bibinfo{author}{\bibfnamefont{C.}~\bibnamefont{Kubel}}, \bibnamefont{and}
  \bibinfo{author}{\bibfnamefont{K.}~\bibnamefont{Mullen}},
  \bibinfo{journal}{Chemistry-A European Journal} \textbf{\bibinfo{volume}{4}},
  \bibinfo{pages}{2099} (\bibinfo{year}{1998}).

\bibitem[{\citenamefont{van~de Craats et~al.}(1999)\citenamefont{van~de Craats,
  Warman, Fechtenkotter, Brand, Harbison, and Mullen}}]{vandeCraatsWFBHM99}
\bibinfo{author}{\bibfnamefont{A.~M.} \bibnamefont{van~de Craats}},
  \bibinfo{author}{\bibfnamefont{J.~M.} \bibnamefont{Warman}},
  \bibinfo{author}{\bibfnamefont{A.}~\bibnamefont{Fechtenkotter}},
  \bibinfo{author}{\bibfnamefont{J.~D.} \bibnamefont{Brand}},
  \bibinfo{author}{\bibfnamefont{M.~A.} \bibnamefont{Harbison}},
  \bibnamefont{and} \bibinfo{author}{\bibfnamefont{K.}~\bibnamefont{Mullen}},
  \bibinfo{journal}{Advanced Materials} \textbf{\bibinfo{volume}{11}},
  \bibinfo{pages}{1469} (\bibinfo{year}{1999}).

\bibitem[{\citenamefont{Schmidt-Mende et~al.}(2001)\citenamefont{Schmidt-Mende,
  Fechtenkotter, Mullen, Moons, Friend, and MacKenzie}}]{Schmidt-MendeFMMFM01}
\bibinfo{author}{\bibfnamefont{L.}~\bibnamefont{Schmidt-Mende}},
  \bibinfo{author}{\bibfnamefont{A.}~\bibnamefont{Fechtenkotter}},
  \bibinfo{author}{\bibfnamefont{K.}~\bibnamefont{Mullen}},
  \bibinfo{author}{\bibfnamefont{E.}~\bibnamefont{Moons}},
  \bibinfo{author}{\bibfnamefont{R.~H.} \bibnamefont{Friend}},
  \bibnamefont{and} \bibinfo{author}{\bibfnamefont{J.~D.}
  \bibnamefont{MacKenzie}}, \bibinfo{journal}{Science}
  \textbf{\bibinfo{volume}{293}}, \bibinfo{pages}{1119} (\bibinfo{year}{2001}).

\bibitem[{\citenamefont{Halls et~al.}(2000)\citenamefont{Halls, Arias,
  MacKenzie, Wu, Inbasekaran, Woo, and Friend}}]{HallsAMWIWF00}
\bibinfo{author}{\bibfnamefont{J.~J.~M.} \bibnamefont{Halls}},
  \bibinfo{author}{\bibfnamefont{A.~C.} \bibnamefont{Arias}},
  \bibinfo{author}{\bibfnamefont{J.~D.} \bibnamefont{MacKenzie}},
  \bibinfo{author}{\bibfnamefont{W.~S.} \bibnamefont{Wu}},
  \bibinfo{author}{\bibfnamefont{M.}~\bibnamefont{Inbasekaran}},
  \bibinfo{author}{\bibfnamefont{E.~P.} \bibnamefont{Woo}}, \bibnamefont{and}
  \bibinfo{author}{\bibfnamefont{R.~H.} \bibnamefont{Friend}},
  \bibinfo{journal}{Advanced Materials} \textbf{\bibinfo{volume}{12}},
  \bibinfo{pages}{498} (\bibinfo{year}{2000}).

\bibitem[{\citenamefont{Shaheen et~al.}(2001)\citenamefont{Shaheen, Brabec,
  Sariciftci, Padinger, Fromherz, and Hummelen}}]{ShaheenBSPFH01}
\bibinfo{author}{\bibfnamefont{S.~E.} \bibnamefont{Shaheen}},
  \bibinfo{author}{\bibfnamefont{C.~J.} \bibnamefont{Brabec}},
  \bibinfo{author}{\bibfnamefont{N.~S.} \bibnamefont{Sariciftci}},
  \bibinfo{author}{\bibfnamefont{F.}~\bibnamefont{Padinger}},
  \bibinfo{author}{\bibfnamefont{T.}~\bibnamefont{Fromherz}}, \bibnamefont{and}
  \bibinfo{author}{\bibfnamefont{J.~C.} \bibnamefont{Hummelen}},
  \bibinfo{journal}{Applied Physics Letters} \textbf{\bibinfo{volume}{78}},
  \bibinfo{pages}{841} (\bibinfo{year}{2001}).

\bibitem[{\citenamefont{Arias et~al.}(2001)\citenamefont{Arias, MacKenzie,
  Stevenson, Halls, Inbasekaran, Woo, Richards, and Friend}}]{AriasMSHIWRF01}
\bibinfo{author}{\bibfnamefont{A.~C.} \bibnamefont{Arias}},
  \bibinfo{author}{\bibfnamefont{J.~D.} \bibnamefont{MacKenzie}},
  \bibinfo{author}{\bibfnamefont{R.}~\bibnamefont{Stevenson}},
  \bibinfo{author}{\bibfnamefont{J.~J.~M.} \bibnamefont{Halls}},
  \bibinfo{author}{\bibfnamefont{M.}~\bibnamefont{Inbasekaran}},
  \bibinfo{author}{\bibfnamefont{E.~P.} \bibnamefont{Woo}},
  \bibinfo{author}{\bibfnamefont{D.}~\bibnamefont{Richards}}, \bibnamefont{and}
  \bibinfo{author}{\bibfnamefont{R.~H.} \bibnamefont{Friend}},
  \bibinfo{journal}{Macromolecules} \textbf{\bibinfo{volume}{34}},
  \bibinfo{pages}{6005} (\bibinfo{year}{2001}).

\bibitem[{\citenamefont{Debije et~al.}(2004)\citenamefont{Debije, Piris,
  de~Haas, Warman, Tomovic, Simpson, Watson, and Mullen}}]{DebijePdWTSWM04}
\bibinfo{author}{\bibfnamefont{M.~G.} \bibnamefont{Debije}},
  \bibinfo{author}{\bibfnamefont{J.}~\bibnamefont{Piris}},
  \bibinfo{author}{\bibfnamefont{M.~P.} \bibnamefont{de~Haas}},
  \bibinfo{author}{\bibfnamefont{J.~M.} \bibnamefont{Warman}},
  \bibinfo{author}{\bibfnamefont{Z.}~\bibnamefont{Tomovic}},
  \bibinfo{author}{\bibfnamefont{C.~D.} \bibnamefont{Simpson}},
  \bibinfo{author}{\bibfnamefont{M.~D.} \bibnamefont{Watson}},
  \bibnamefont{and} \bibinfo{author}{\bibfnamefont{K.}~\bibnamefont{Mullen}},
  \bibinfo{journal}{J. Am. Chem. Soc.} \textbf{\bibinfo{volume}{126}},
  \bibinfo{pages}{4641} (\bibinfo{year}{2004}).

\bibitem[{\citenamefont{Iftimie et~al.}(2005)\citenamefont{Iftimie, Minary, and
  Tuckerman}}]{AIMD_PNAS_TUCKERMAN2005}
\bibinfo{author}{\bibfnamefont{R.}~\bibnamefont{Iftimie}},
  \bibinfo{author}{\bibfnamefont{P.}~\bibnamefont{Minary}}, \bibnamefont{and}
  \bibinfo{author}{\bibfnamefont{M.~E.} \bibnamefont{Tuckerman}},
  \bibinfo{journal}{Proc. Natl Acad. Sci. USA} \textbf{\bibinfo{volume}{102}},
  \bibinfo{pages}{6654} (\bibinfo{year}{2005}).

\bibitem[{\citenamefont{von Lilienfeld
  et~al.}(2005{\natexlab{a}})\citenamefont{von Lilienfeld, Lins, and
  Rothlisberger}}]{myself-prl2005}
\bibinfo{author}{\bibfnamefont{O.~A.} \bibnamefont{von Lilienfeld}},
  \bibinfo{author}{\bibfnamefont{R.}~\bibnamefont{Lins}}, \bibnamefont{and}
  \bibinfo{author}{\bibfnamefont{U.}~\bibnamefont{Rothlisberger}},
  \bibinfo{journal}{Phys. Rev. Lett.} \textbf{\bibinfo{volume}{95}},
  \bibinfo{pages}{153002} (\bibinfo{year}{2005}{\natexlab{a}}).

\bibitem[{\citenamefont{Maliniak}(1992)}]{Maliniak92}
\bibinfo{author}{\bibfnamefont{A.}~\bibnamefont{Maliniak}},
  \bibinfo{journal}{Journal of Chemical Physics} \textbf{\bibinfo{volume}{96}},
  \bibinfo{pages}{2306} (\bibinfo{year}{1992}).

\bibitem[{\citenamefont{Ono and Kondo}(1992)}]{OnoK92}
\bibinfo{author}{\bibfnamefont{I.}~\bibnamefont{Ono}} \bibnamefont{and}
  \bibinfo{author}{\bibfnamefont{S.}~\bibnamefont{Kondo}},
  \bibinfo{journal}{Bulletin of The Chemical Society of Japan}
  \textbf{\bibinfo{volume}{65}}, \bibinfo{pages}{1057} (\bibinfo{year}{1992}).

\bibitem[{\citenamefont{Mulder et~al.}(2003)\citenamefont{Mulder, Stride,
  Picken, Kouwer, de~Haas, Siebbeles, and Kearley}}]{MulderSPKdSK03}
\bibinfo{author}{\bibfnamefont{F.~M.} \bibnamefont{Mulder}},
  \bibinfo{author}{\bibfnamefont{J.}~\bibnamefont{Stride}},
  \bibinfo{author}{\bibfnamefont{S.~J.} \bibnamefont{Picken}},
  \bibinfo{author}{\bibfnamefont{P.~H.~J.} \bibnamefont{Kouwer}},
  \bibinfo{author}{\bibfnamefont{M.~P.} \bibnamefont{de~Haas}},
  \bibinfo{author}{\bibfnamefont{L.~D.~A.} \bibnamefont{Siebbeles}},
  \bibnamefont{and} \bibinfo{author}{\bibfnamefont{G.~J.}
  \bibnamefont{Kearley}}, \bibinfo{journal}{Journal of The American Chemical
  Society} \textbf{\bibinfo{volume}{125}}, \bibinfo{pages}{3860}
  (\bibinfo{year}{2003}).

\bibitem[{\citenamefont{Cinacchi et~al.}(2004)\citenamefont{Cinacchi, Colle,
  and Tani}}]{CinacchiCT04}
\bibinfo{author}{\bibfnamefont{G.}~\bibnamefont{Cinacchi}},
  \bibinfo{author}{\bibfnamefont{R.}~\bibnamefont{Colle}}, \bibnamefont{and}
  \bibinfo{author}{\bibfnamefont{A.}~\bibnamefont{Tani}},
  \bibinfo{journal}{Journal of Physical Chemistry B}
  \textbf{\bibinfo{volume}{108}}, \bibinfo{pages}{7969} (\bibinfo{year}{2004}).

\bibitem[{\citenamefont{Veerman and Frenkel}(1992)}]{VeermanF92}
\bibinfo{author}{\bibfnamefont{J.~A.~C.} \bibnamefont{Veerman}}
  \bibnamefont{and} \bibinfo{author}{\bibfnamefont{D.}~\bibnamefont{Frenkel}},
  \bibinfo{journal}{Physical Review A} \textbf{\bibinfo{volume}{45}},
  \bibinfo{pages}{5632} (\bibinfo{year}{1992}).

\bibitem[{\citenamefont{Emerson et~al.}(1994)\citenamefont{Emerson, Luckhurst,
  and Whatling}}]{EmersonLW94}
\bibinfo{author}{\bibfnamefont{A.~P.~J.} \bibnamefont{Emerson}},
  \bibinfo{author}{\bibfnamefont{G.~R.} \bibnamefont{Luckhurst}},
  \bibnamefont{and} \bibinfo{author}{\bibfnamefont{S.~G.}
  \bibnamefont{Whatling}}, \bibinfo{journal}{Molecular Physics}
  \textbf{\bibinfo{volume}{82}}, \bibinfo{pages}{113} (\bibinfo{year}{1994}).

\bibitem[{\citenamefont{Bates and Luckhurst}(1996)}]{BatesL96}
\bibinfo{author}{\bibfnamefont{M.~A.} \bibnamefont{Bates}} \bibnamefont{and}
  \bibinfo{author}{\bibfnamefont{G.~R.} \bibnamefont{Luckhurst}},
  \bibinfo{journal}{Journal of Chemical Physics}
  \textbf{\bibinfo{volume}{104}}, \bibinfo{pages}{6696} (\bibinfo{year}{1996}).

\bibitem[{\citenamefont{Zewdie}(1998)}]{Zewdie98}
\bibinfo{author}{\bibfnamefont{H.}~\bibnamefont{Zewdie}},
  \bibinfo{journal}{Physical Review E} \textbf{\bibinfo{volume}{57}},
  \bibinfo{pages}{1793} (\bibinfo{year}{1998}).

\bibitem[{\citenamefont{Cinacchi and Tani}(2002)}]{CinacchiT02}
\bibinfo{author}{\bibfnamefont{G.}~\bibnamefont{Cinacchi}} \bibnamefont{and}
  \bibinfo{author}{\bibfnamefont{A.}~\bibnamefont{Tani}},
  \bibinfo{journal}{Journal of Chemical Physics}
  \textbf{\bibinfo{volume}{117}}, \bibinfo{pages}{11388}
  (\bibinfo{year}{2002}).

\bibitem[{\citenamefont{Caprion et~al.}(2003)\citenamefont{Caprion,
  Bellier-Castella, and Ryckaert}}]{CaprionBR03}
\bibinfo{author}{\bibfnamefont{D.}~\bibnamefont{Caprion}},
  \bibinfo{author}{\bibfnamefont{L.}~\bibnamefont{Bellier-Castella}},
  \bibnamefont{and} \bibinfo{author}{\bibfnamefont{J.~P.}
  \bibnamefont{Ryckaert}}, \bibinfo{journal}{Physical Review E}
  \textbf{\bibinfo{volume}{67}} (\bibinfo{year}{2003}).

\bibitem[{\citenamefont{Bellier-Castella
  et~al.}(2004)\citenamefont{Bellier-Castella, Caprion, and
  Ryckaert}}]{Bellier-CastellaCR04}
\bibinfo{author}{\bibfnamefont{L.}~\bibnamefont{Bellier-Castella}},
  \bibinfo{author}{\bibfnamefont{D.}~\bibnamefont{Caprion}}, \bibnamefont{and}
  \bibinfo{author}{\bibfnamefont{J.~P.} \bibnamefont{Ryckaert}},
  \bibinfo{journal}{Journal of Chemical Physics}
  \textbf{\bibinfo{volume}{121}}, \bibinfo{pages}{4874} (\bibinfo{year}{2004}).

\bibitem[{\citenamefont{Baschnagel et~al.}(2000)\citenamefont{Baschnagel,
  Binder, Doruker, Gusev, Hahn, Kremer, Mattice, Muller-Plathe, Murat, Paul
  et~al.}}]{Baschnagel00}
\bibinfo{author}{\bibfnamefont{J.}~\bibnamefont{Baschnagel}},
  \bibinfo{author}{\bibfnamefont{K.}~\bibnamefont{Binder}},
  \bibinfo{author}{\bibfnamefont{P.}~\bibnamefont{Doruker}},
  \bibinfo{author}{\bibfnamefont{A.~A.} \bibnamefont{Gusev}},
  \bibinfo{author}{\bibfnamefont{O.}~\bibnamefont{Hahn}},
  \bibinfo{author}{\bibfnamefont{K.}~\bibnamefont{Kremer}},
  \bibinfo{author}{\bibfnamefont{W.~L.} \bibnamefont{Mattice}},
  \bibinfo{author}{\bibfnamefont{F.}~\bibnamefont{Muller-Plathe}},
  \bibinfo{author}{\bibfnamefont{M.}~\bibnamefont{Murat}},
  \bibinfo{author}{\bibfnamefont{W.}~\bibnamefont{Paul}}, \bibnamefont{et~al.},
  \bibinfo{journal}{Advances In Polymer Science: Viscoelasticity, Atomistic
  Models, Statistical Chemistry} \textbf{\bibinfo{volume}{152}},
  \bibinfo{pages}{41} (\bibinfo{year}{2000}).

\bibitem[{\citenamefont{Abrams et~al.}(2003)\citenamefont{Abrams, {Delle Site},
  and Kremer}}]{AbramsDK03}
\bibinfo{author}{\bibfnamefont{C.~F.} \bibnamefont{Abrams}},
  \bibinfo{author}{\bibfnamefont{L.}~\bibnamefont{{Delle Site}}},
  \bibnamefont{and} \bibinfo{author}{\bibfnamefont{K.}~\bibnamefont{Kremer}},
  \bibinfo{journal}{Phys. Rev. E} \textbf{\bibinfo{volume}{67}},
  \bibinfo{pages}{021807} (\bibinfo{year}{2003}).

\bibitem[{\citenamefont{{Delle Site} et~al.}(2002)\citenamefont{{Delle Site},
  Abrams, Alavi, and Kremer}}]{DelleSiteAAK02}
\bibinfo{author}{\bibfnamefont{L.}~\bibnamefont{{Delle Site}}},
  \bibinfo{author}{\bibfnamefont{C.~F.} \bibnamefont{Abrams}},
  \bibinfo{author}{\bibfnamefont{A.}~\bibnamefont{Alavi}}, \bibnamefont{and}
  \bibinfo{author}{\bibfnamefont{K.}~\bibnamefont{Kremer}},
  \bibinfo{journal}{Phys. Rev. Lett.} \textbf{\bibinfo{volume}{89}},
  \bibinfo{pages}{156103} (\bibinfo{year}{2002}).

\bibitem[{\citenamefont{Abrams and Kremer}(2003)}]{AbramsK03}
\bibinfo{author}{\bibfnamefont{C.~F.} \bibnamefont{Abrams}} \bibnamefont{and}
  \bibinfo{author}{\bibfnamefont{K.}~\bibnamefont{Kremer}},
  \bibinfo{journal}{Macromolecules} \textbf{\bibinfo{volume}{36}},
  \bibinfo{pages}{260} (\bibinfo{year}{2003}).

\bibitem[{\citenamefont{Zhou et~al.}(2005{\natexlab{a}})\citenamefont{Zhou,
  Andrienko, {Delle Site}, and Kremer}}]{zhou2005a}
\bibinfo{author}{\bibfnamefont{X.}~\bibnamefont{Zhou}},
  \bibinfo{author}{\bibfnamefont{D.}~\bibnamefont{Andrienko}},
  \bibinfo{author}{\bibfnamefont{L.}~\bibnamefont{{Delle Site}}},
  \bibnamefont{and} \bibinfo{author}{\bibfnamefont{K.}~\bibnamefont{Kremer}},
  \bibinfo{journal}{Europhys. Lett.} \textbf{\bibinfo{volume}{70}},
  \bibinfo{pages}{264} (\bibinfo{year}{2005}{\natexlab{a}}).

\bibitem[{\citenamefont{Zhou et~al.}(2005{\natexlab{b}})\citenamefont{Zhou,
  Andrienko, {Delle Site}, and Kremer}}]{zhou:2005.b}
\bibinfo{author}{\bibfnamefont{X.}~\bibnamefont{Zhou}},
  \bibinfo{author}{\bibfnamefont{D.}~\bibnamefont{Andrienko}},
  \bibinfo{author}{\bibfnamefont{L.}~\bibnamefont{{Delle Site}}},
  \bibnamefont{and} \bibinfo{author}{\bibfnamefont{K.}~\bibnamefont{Kremer}},
  \bibinfo{journal}{J. Chem. Phys.} \textbf{\bibinfo{volume}{123}},
  \bibinfo{pages}{104904} (\bibinfo{year}{2005}{\natexlab{b}}).

\bibitem[{\citenamefont{Andrienko et~al.}(2005)\citenamefont{Andrienko, Leon,
  Site, and Kremer}}]{andrienko:2005.b}
\bibinfo{author}{\bibfnamefont{D.}~\bibnamefont{Andrienko}},
  \bibinfo{author}{\bibfnamefont{S.}~\bibnamefont{Leon}},
  \bibinfo{author}{\bibfnamefont{L.~D.} \bibnamefont{Site}}, \bibnamefont{and}
  \bibinfo{author}{\bibfnamefont{K.}~\bibnamefont{Kremer}},
  \bibinfo{journal}{Macromolecules} \textbf{\bibinfo{volume}{38}},
  \bibinfo{pages}{5810 } (\bibinfo{year}{2005}).

\bibitem[{\citenamefont{Craig and Thirunamachandran}(1998)}]{MQED_CRAIG}
\bibinfo{author}{\bibfnamefont{D.~P.} \bibnamefont{Craig}} \bibnamefont{and}
  \bibinfo{author}{\bibfnamefont{T.}~\bibnamefont{Thirunamachandran}}, in
  \emph{\bibinfo{booktitle}{Molecular Quantum Electrodynamics}}
  (\bibinfo{publisher}{Dover Publications, Inc., Mineola, New York},
  \bibinfo{year}{1998}).

\bibitem[{\citenamefont{Tsuzuki et~al.}(2002)\citenamefont{Tsuzuki, Uchimaru,
  Sugawara, and Mikami}}]{TsuzukisBenzeneDimer}
\bibinfo{author}{\bibfnamefont{S.}~\bibnamefont{Tsuzuki}},
  \bibinfo{author}{\bibfnamefont{T.}~\bibnamefont{Uchimaru}},
  \bibinfo{author}{\bibfnamefont{K.}~\bibnamefont{Sugawara}}, \bibnamefont{and}
  \bibinfo{author}{\bibfnamefont{M.}~\bibnamefont{Mikami}},
  \bibinfo{journal}{J. Chem. Phys.} \textbf{\bibinfo{volume}{117}},
  \bibinfo{pages}{11216} (\bibinfo{year}{2002}).

\bibitem[{\citenamefont{Koch and Holthausen}(2002)}]{ChemistsGuidetoDFT}
\bibinfo{author}{\bibfnamefont{W.}~\bibnamefont{Koch}} \bibnamefont{and}
  \bibinfo{author}{\bibfnamefont{M.~C.} \bibnamefont{Holthausen}},
  \emph{\bibinfo{title}{A Chemist's Guide to Density Functional Theory}}
  (\bibinfo{publisher}{Wiley-VCH}, \bibinfo{year}{2002}).

\bibitem[{\citenamefont{Kristy\'an and Pulay}(1994)}]{Pulay-NoVdw}
\bibinfo{author}{\bibfnamefont{S.}~\bibnamefont{Kristy\'an}} \bibnamefont{and}
  \bibinfo{author}{\bibfnamefont{P.}~\bibnamefont{Pulay}},
  \bibinfo{journal}{Chem. Phys. Lett.} \textbf{\bibinfo{volume}{229}},
  \bibinfo{pages}{175} (\bibinfo{year}{1994}).

\bibitem[{\citenamefont{P\'erez-Jord\'a and Becke}(1995)}]{Becke-NoVdw}
\bibinfo{author}{\bibfnamefont{J.~M.} \bibnamefont{P\'erez-Jord\'a}}
  \bibnamefont{and} \bibinfo{author}{\bibfnamefont{A.~D.} \bibnamefont{Becke}},
  \bibinfo{journal}{Chem. Phys. Lett.} \textbf{\bibinfo{volume}{233}},
  \bibinfo{pages}{134} (\bibinfo{year}{1995}).

\bibitem[{\citenamefont{Meijer and Sprik}(1996)}]{Sprik-NoVdw}
\bibinfo{author}{\bibfnamefont{E.~J.} \bibnamefont{Meijer}} \bibnamefont{and}
  \bibinfo{author}{\bibfnamefont{M.}~\bibnamefont{Sprik}}, \bibinfo{journal}{J.
  Chem. Phys.} \textbf{\bibinfo{volume}{105}}, \bibinfo{pages}{8684}
  (\bibinfo{year}{1996}).

\bibitem[{\citenamefont{van Mourik and Gdanitz}(2002)}]{CriticalNote_DFT4He}
\bibinfo{author}{\bibfnamefont{T.}~\bibnamefont{van Mourik}} \bibnamefont{and}
  \bibinfo{author}{\bibfnamefont{R.~J.} \bibnamefont{Gdanitz}},
  \bibinfo{journal}{J. Chem. Phys.} \textbf{\bibinfo{volume}{116}},
  \bibinfo{pages}{9620} (\bibinfo{year}{2002}).

\bibitem[{\citenamefont{Wu et~al.}(2001)\citenamefont{Wu, Vargas, Nayak,
  Lotrich, and Scoles}}]{ScolesVDW}
\bibinfo{author}{\bibfnamefont{X.}~\bibnamefont{Wu}},
  \bibinfo{author}{\bibfnamefont{M.~C.} \bibnamefont{Vargas}},
  \bibinfo{author}{\bibfnamefont{S.}~\bibnamefont{Nayak}},
  \bibinfo{author}{\bibfnamefont{V.}~\bibnamefont{Lotrich}}, \bibnamefont{and}
  \bibinfo{author}{\bibfnamefont{G.}~\bibnamefont{Scoles}},
  \bibinfo{journal}{J. Chem. Phys.} \textbf{\bibinfo{volume}{115}},
  \bibinfo{pages}{8748} (\bibinfo{year}{2001}).

\bibitem[{\citenamefont{Wesolowski et~al.}(2002)\citenamefont{Wesolowski,
  Morgantini, and Weber}}]{Wesolowski-VDW1}
\bibinfo{author}{\bibfnamefont{T.~A.} \bibnamefont{Wesolowski}},
  \bibinfo{author}{\bibfnamefont{P.~Y.} \bibnamefont{Morgantini}},
  \bibnamefont{and} \bibinfo{author}{\bibfnamefont{J.}~\bibnamefont{Weber}},
  \bibinfo{journal}{J. Chem. Phys.} \textbf{\bibinfo{volume}{116}},
  \bibinfo{pages}{6411} (\bibinfo{year}{2002}).

\bibitem[{\citenamefont{Wesolowski and Tran}(2003)}]{Wesolowski-VDW2}
\bibinfo{author}{\bibfnamefont{T.~A.} \bibnamefont{Wesolowski}}
  \bibnamefont{and} \bibinfo{author}{\bibfnamefont{F.}~\bibnamefont{Tran}},
  \bibinfo{journal}{J. Chem. Phys.} \textbf{\bibinfo{volume}{118}},
  \bibinfo{pages}{2072} (\bibinfo{year}{2003}).

\bibitem[{\citenamefont{Andersson et~al.}(1996)\citenamefont{Andersson,
  Langreth, and Lundqvist}}]{nonlocalVDW1}
\bibinfo{author}{\bibfnamefont{Y.}~\bibnamefont{Andersson}},
  \bibinfo{author}{\bibfnamefont{D.~C.} \bibnamefont{Langreth}},
  \bibnamefont{and} \bibinfo{author}{\bibfnamefont{B.~I.}
  \bibnamefont{Lundqvist}}, \bibinfo{journal}{Phys. Rev. Lett.}
  \textbf{\bibinfo{volume}{76}}, \bibinfo{pages}{102} (\bibinfo{year}{1996}).

\bibitem[{\citenamefont{Hult et~al.}(1996)\citenamefont{Hult, Andersson, and
  Lundqvist}}]{nonlocalVDW2}
\bibinfo{author}{\bibfnamefont{E.}~\bibnamefont{Hult}},
  \bibinfo{author}{\bibfnamefont{Y.}~\bibnamefont{Andersson}},
  \bibnamefont{and} \bibinfo{author}{\bibfnamefont{B.~I.}
  \bibnamefont{Lundqvist}}, \bibinfo{journal}{Phys. Rev. Lett.}
  \textbf{\bibinfo{volume}{77}}, \bibinfo{pages}{2029} (\bibinfo{year}{1996}).

\bibitem[{\citenamefont{Rydberg et~al.}(2000)\citenamefont{Rydberg, Lundqvist,
  Langreth, and Dion}}]{nonlocalVDW3}
\bibinfo{author}{\bibfnamefont{H.}~\bibnamefont{Rydberg}},
  \bibinfo{author}{\bibfnamefont{B.~I.} \bibnamefont{Lundqvist}},
  \bibinfo{author}{\bibfnamefont{D.~C.} \bibnamefont{Langreth}},
  \bibnamefont{and} \bibinfo{author}{\bibfnamefont{M.}~\bibnamefont{Dion}},
  \bibinfo{journal}{Phys. Rev. B} \textbf{\bibinfo{volume}{62}},
  \bibinfo{pages}{6997} (\bibinfo{year}{2000}).

\bibitem[{\citenamefont{Rydberg et~al.}(2003)\citenamefont{Rydberg, Dion,
  Jacobsen, Schr\"oder, Hyldgaard, Simak, Langreth, and
  Lundqvist}}]{nonlocalVDW4}
\bibinfo{author}{\bibfnamefont{H.}~\bibnamefont{Rydberg}},
  \bibinfo{author}{\bibfnamefont{M.}~\bibnamefont{Dion}},
  \bibinfo{author}{\bibfnamefont{N.}~\bibnamefont{Jacobsen}},
  \bibinfo{author}{\bibfnamefont{E.}~\bibnamefont{Schr\"oder}},
  \bibinfo{author}{\bibfnamefont{P.}~\bibnamefont{Hyldgaard}},
  \bibinfo{author}{\bibfnamefont{S.~I.} \bibnamefont{Simak}},
  \bibinfo{author}{\bibfnamefont{D.~C.} \bibnamefont{Langreth}},
  \bibnamefont{and} \bibinfo{author}{\bibfnamefont{B.~I.}
  \bibnamefont{Lundqvist}}, \bibinfo{journal}{Phys. Rev. Lett.}
  \textbf{\bibinfo{volume}{91}}, \bibinfo{pages}{126402}
  (\bibinfo{year}{2003}).

\bibitem[{\citenamefont{Dion et~al.}(2004)\citenamefont{Dion, Rydberg,
  Schr\"oder, Langreth, and Lundqvist}}]{nonlocalVDW5}
\bibinfo{author}{\bibfnamefont{M.}~\bibnamefont{Dion}},
  \bibinfo{author}{\bibfnamefont{H.}~\bibnamefont{Rydberg}},
  \bibinfo{author}{\bibfnamefont{E.}~\bibnamefont{Schr\"oder}},
  \bibinfo{author}{\bibfnamefont{D.~C.} \bibnamefont{Langreth}},
  \bibnamefont{and} \bibinfo{author}{\bibfnamefont{B.~I.}
  \bibnamefont{Lundqvist}}, \bibinfo{journal}{Phys. Rev. Lett.}
  \textbf{\bibinfo{volume}{92}}, \bibinfo{pages}{246401}
  (\bibinfo{year}{2004}).

\bibitem[{\citenamefont{Misquitta et~al.}(2003)\citenamefont{Misquitta,
  Jeziorski, and Szalewicz}}]{SzalewiczVDW}
\bibinfo{author}{\bibfnamefont{A.~J.} \bibnamefont{Misquitta}},
  \bibinfo{author}{\bibfnamefont{B.}~\bibnamefont{Jeziorski}},
  \bibnamefont{and}
  \bibinfo{author}{\bibfnamefont{K.}~\bibnamefont{Szalewicz}},
  \bibinfo{journal}{Phys. Rev. Lett.} \textbf{\bibinfo{volume}{91}},
  \bibinfo{pages}{33201} (\bibinfo{year}{2003}).

\bibitem[{\citenamefont{Kohn et~al.}(1998)\citenamefont{Kohn, Meir, and
  Makarov}}]{KohnVdwDFT}
\bibinfo{author}{\bibfnamefont{W.}~\bibnamefont{Kohn}},
  \bibinfo{author}{\bibfnamefont{Y.}~\bibnamefont{Meir}}, \bibnamefont{and}
  \bibinfo{author}{\bibfnamefont{D.~E.} \bibnamefont{Makarov}},
  \bibinfo{journal}{Phys. Rev. Lett.} \textbf{\bibinfo{volume}{80}},
  \bibinfo{pages}{4153} (\bibinfo{year}{1998}).

\bibitem[{\citenamefont{LeSar}(1984)}]{LeSar-VDW}
\bibinfo{author}{\bibfnamefont{R.}~\bibnamefont{LeSar}}, \bibinfo{journal}{J.
  Phys. Chem.} \textbf{\bibinfo{volume}{88}}, \bibinfo{pages}{4272}
  (\bibinfo{year}{1984}).

\bibitem[{\citenamefont{Grimme}(2004)}]{Grimme2004}
\bibinfo{author}{\bibfnamefont{S.}~\bibnamefont{Grimme}}, \bibinfo{journal}{J.
  Comp. Chem.} \textbf{\bibinfo{volume}{25}}, \bibinfo{pages}{1463}
  (\bibinfo{year}{2004}).

\bibitem[{\citenamefont{van Gisbergen et~al.}(1995)\citenamefont{van Gisbergen,
  Snijders, and Baerends}}]{BaerendsC6}
\bibinfo{author}{\bibfnamefont{S.~J.~A.} \bibnamefont{van Gisbergen}},
  \bibinfo{author}{\bibfnamefont{J.~G.} \bibnamefont{Snijders}},
  \bibnamefont{and} \bibinfo{author}{\bibfnamefont{E.~J.}
  \bibnamefont{Baerends}}, \bibinfo{journal}{J. Chem. Phys.}
  \textbf{\bibinfo{volume}{103}}, \bibinfo{pages}{9347} (\bibinfo{year}{1995}).

\bibitem[{\citenamefont{Johnson and Becke}(2005)}]{BeckesC6}
\bibinfo{author}{\bibfnamefont{E.~R.} \bibnamefont{Johnson}} \bibnamefont{and}
  \bibinfo{author}{\bibfnamefont{A.~D.} \bibnamefont{Becke}},
  \bibinfo{journal}{J. Chem. Phys.} \textbf{\bibinfo{volume}{123}},
  \bibinfo{pages}{24101} (\bibinfo{year}{2005}).

\bibitem[{\citenamefont{Ortmann et~al.}(2005)\citenamefont{Ortmann, Schmidt,
  and Bechstedt}}]{BechstedtPRL2005}
\bibinfo{author}{\bibfnamefont{F.}~\bibnamefont{Ortmann}},
  \bibinfo{author}{\bibfnamefont{W.~G.} \bibnamefont{Schmidt}},
  \bibnamefont{and}
  \bibinfo{author}{\bibfnamefont{F.}~\bibnamefont{Bechstedt}},
  \bibinfo{journal}{Phys. Rev. Lett.} \textbf{\bibinfo{volume}{95}},
  \bibinfo{pages}{186101} (\bibinfo{year}{2005}).

\bibitem[{\citenamefont{von Lilienfeld et~al.}(2004)\citenamefont{von
  Lilienfeld, Tavernelli, Rothlisberger, and Sebastiani}}]{myself-prl2004}
\bibinfo{author}{\bibfnamefont{O.~A.} \bibnamefont{von Lilienfeld}},
  \bibinfo{author}{\bibfnamefont{I.}~\bibnamefont{Tavernelli}},
  \bibinfo{author}{\bibfnamefont{U.}~\bibnamefont{Rothlisberger}},
  \bibnamefont{and}
  \bibinfo{author}{\bibfnamefont{D.}~\bibnamefont{Sebastiani}},
  \bibinfo{journal}{Phys. Rev. Lett.} \textbf{\bibinfo{volume}{93}},
  \bibinfo{pages}{153004} (\bibinfo{year}{2004}).

\bibitem[{\citenamefont{von Lilienfeld
  et~al.}(2005{\natexlab{b}})\citenamefont{von Lilienfeld, Tavernelli,
  Rothlisberger, and Sebastiani}}]{myself-jcp2004}
\bibinfo{author}{\bibfnamefont{O.~A.} \bibnamefont{von Lilienfeld}},
  \bibinfo{author}{\bibfnamefont{I.}~\bibnamefont{Tavernelli}},
  \bibinfo{author}{\bibfnamefont{U.}~\bibnamefont{Rothlisberger}},
  \bibnamefont{and}
  \bibinfo{author}{\bibfnamefont{D.}~\bibnamefont{Sebastiani}},
  \bibinfo{journal}{J. Chem. Phys.} \textbf{\bibinfo{volume}{122}},
  \bibinfo{pages}{14113} (\bibinfo{year}{2005}{\natexlab{b}}).

\bibitem[{\citenamefont{von Lilienfeld
  et~al.}(2005{\natexlab{c}})\citenamefont{von Lilienfeld, Tavernelli,
  Rothlisberger, and Sebastiani}}]{myself-prb2005}
\bibinfo{author}{\bibfnamefont{O.~A.} \bibnamefont{von Lilienfeld}},
  \bibinfo{author}{\bibfnamefont{I.}~\bibnamefont{Tavernelli}},
  \bibinfo{author}{\bibfnamefont{U.}~\bibnamefont{Rothlisberger}},
  \bibnamefont{and}
  \bibinfo{author}{\bibfnamefont{D.}~\bibnamefont{Sebastiani}},
  \bibinfo{journal}{Phys. Rev. B} \textbf{\bibinfo{volume}{71}},
  \bibinfo{pages}{195119} (\bibinfo{year}{2005}{\natexlab{c}}).

\bibitem[{\citenamefont{Tapavicza et~al.}(2005)\citenamefont{Tapavicza, von
  Lilienfeld, Lin, Coutinho, Tavernelli, and
  Rothlisberger}}]{myself-enrico2005}
\bibinfo{author}{\bibfnamefont{E.}~\bibnamefont{Tapavicza}},
  \bibinfo{author}{\bibfnamefont{O.~A.} \bibnamefont{von Lilienfeld}},
  \bibinfo{author}{\bibfnamefont{I.}~\bibnamefont{Lin}},
  \bibinfo{author}{\bibfnamefont{M.}~\bibnamefont{Coutinho}},
  \bibinfo{author}{\bibfnamefont{I.}~\bibnamefont{Tavernelli}},
  \bibnamefont{and}
  \bibinfo{author}{\bibfnamefont{U.}~\bibnamefont{Rothlisberger}}
  (\bibinfo{year}{2005}), \bibinfo{note}{unpublished}.

\bibitem[{\citenamefont{Coutinho et~al.}(2005)\citenamefont{Coutinho, Lin, von
  Lilienfeld, Tavernelli, and Rothlisberger}}]{myself-mauricio2006}
\bibinfo{author}{\bibfnamefont{M.}~\bibnamefont{Coutinho}},
  \bibinfo{author}{\bibfnamefont{I.}~\bibnamefont{Lin}},
  \bibinfo{author}{\bibfnamefont{O.~A.} \bibnamefont{von Lilienfeld}},
  \bibinfo{author}{\bibfnamefont{I.}~\bibnamefont{Tavernelli}},
  \bibnamefont{and}
  \bibinfo{author}{\bibfnamefont{U.}~\bibnamefont{Rothlisberger}}
  (\bibinfo{year}{2005}), \bibinfo{note}{unpublished}.

\bibitem[{\citenamefont{{J. Hutter et al.}}()}]{cpmd3.92}
\bibinfo{author}{\bibnamefont{{J. Hutter et al.}}},
  \emph{\bibinfo{title}{{Computer code \textsc{CPMD}, version 3.92}}},
  \bibinfo{note}{{Copyright IBM Corp.\ 1990-2001, Copyright MPI-FKF Stuttgart
  1997-2004, \texttt{http://www.cpmd.org}}}.

\bibitem[{\citenamefont{Becke}(1988)}]{B88X}
\bibinfo{author}{\bibfnamefont{A.~D.} \bibnamefont{Becke}},
  \bibinfo{journal}{Phys. Rev. A} \textbf{\bibinfo{volume}{38}},
  \bibinfo{pages}{3098} (\bibinfo{year}{1988}).

\bibitem[{\citenamefont{Colle and Salvetti}(1975)}]{ColleSalvetti}
\bibinfo{author}{\bibfnamefont{R.}~\bibnamefont{Colle}} \bibnamefont{and}
  \bibinfo{author}{\bibfnamefont{D.}~\bibnamefont{Salvetti}},
  \bibinfo{journal}{Theor. Chim. Acta} \textbf{\bibinfo{volume}{37}},
  \bibinfo{pages}{329} (\bibinfo{year}{1975}).

\bibitem[{\citenamefont{Lee et~al.}(1988)\citenamefont{Lee, Yang, and
  Parr}}]{lyp}
\bibinfo{author}{\bibfnamefont{C.}~\bibnamefont{Lee}},
  \bibinfo{author}{\bibfnamefont{W.}~\bibnamefont{Yang}}, \bibnamefont{and}
  \bibinfo{author}{\bibfnamefont{R.~G.} \bibnamefont{Parr}},
  \bibinfo{journal}{Phys. Rev. B} \textbf{\bibinfo{volume}{37}},
  \bibinfo{pages}{785} (\bibinfo{year}{1988}).

\bibitem[{\citenamefont{Perdew}(1986)}]{perdew-BP}
\bibinfo{author}{\bibfnamefont{J.~P.} \bibnamefont{Perdew}},
  \bibinfo{journal}{Phys. Rev. B} \textbf{\bibinfo{volume}{33}},
  \bibinfo{pages}{8822} (\bibinfo{year}{1986}).

\bibitem[{\citenamefont{Perdew et~al.}(1996)\citenamefont{Perdew, Burke, and
  Ernzerhof}}]{PBE}
\bibinfo{author}{\bibfnamefont{J.~P.} \bibnamefont{Perdew}},
  \bibinfo{author}{\bibfnamefont{K.}~\bibnamefont{Burke}}, \bibnamefont{and}
  \bibinfo{author}{\bibfnamefont{M.}~\bibnamefont{Ernzerhof}},
  \bibinfo{journal}{Phys. Rev. Lett.} \textbf{\bibinfo{volume}{77}},
  \bibinfo{pages}{3865} (\bibinfo{year}{1996}).

\bibitem[{\citenamefont{Perdew and Zunger}(1981)}]{perdew-xc}
\bibinfo{author}{\bibfnamefont{J.~P.} \bibnamefont{Perdew}} \bibnamefont{and}
  \bibinfo{author}{\bibfnamefont{A.}~\bibnamefont{Zunger}},
  \bibinfo{journal}{Phys. Rev. B} \textbf{\bibinfo{volume}{23}},
  \bibinfo{pages}{5048} (\bibinfo{year}{1981}).

\bibitem[{\citenamefont{Ceperley and Alder}(1980)}]{ceperley-xc}
\bibinfo{author}{\bibfnamefont{D.~M.} \bibnamefont{Ceperley}} \bibnamefont{and}
  \bibinfo{author}{\bibfnamefont{B.~J.} \bibnamefont{Alder}},
  \bibinfo{journal}{Phys. Rev. Lett.} \textbf{\bibinfo{volume}{45}},
  \bibinfo{pages}{566} (\bibinfo{year}{1980}).

\bibitem[{\citenamefont{Goedecker et~al.}(1996)\citenamefont{Goedecker, Teter,
  and Hutter}}]{SG}
\bibinfo{author}{\bibfnamefont{S.}~\bibnamefont{Goedecker}},
  \bibinfo{author}{\bibfnamefont{M.}~\bibnamefont{Teter}}, \bibnamefont{and}
  \bibinfo{author}{\bibfnamefont{J.}~\bibnamefont{Hutter}},
  \bibinfo{journal}{Phys. Rev. B} \textbf{\bibinfo{volume}{54}},
  \bibinfo{pages}{1703} (\bibinfo{year}{1996}).

\bibitem[{\citenamefont{Hartwigsen et~al.}(1998)\citenamefont{Hartwigsen,
  Goedecker, and Hutter}}]{sgpsp}
\bibinfo{author}{\bibfnamefont{C.}~\bibnamefont{Hartwigsen}},
  \bibinfo{author}{\bibfnamefont{S.}~\bibnamefont{Goedecker}},
  \bibnamefont{and} \bibinfo{author}{\bibfnamefont{J.}~\bibnamefont{Hutter}},
  \bibinfo{journal}{Phys. Rev. B} \textbf{\bibinfo{volume}{58}},
  \bibinfo{pages}{3641} (\bibinfo{year}{1998}).

\bibitem[{\citenamefont{Krack}(2005)}]{krackPP}
\bibinfo{author}{\bibfnamefont{M.}~\bibnamefont{Krack}},
  \bibinfo{journal}{Theor. Chim. Acta} \textbf{\bibinfo{volume}{114}},
  \bibinfo{pages}{145} (\bibinfo{year}{2005}).

\bibitem[{\citenamefont{Martyna and Tuckerman}(1999)}]{martyna-tuckerman}
\bibinfo{author}{\bibfnamefont{G.}~\bibnamefont{Martyna}} \bibnamefont{and}
  \bibinfo{author}{\bibfnamefont{M.}~\bibnamefont{Tuckerman}},
  \bibinfo{journal}{J. Chem. Phys.} \textbf{\bibinfo{volume}{110}},
  \bibinfo{pages}{2810} (\bibinfo{year}{1999}).

\bibitem[{\citenamefont{Zacharia et~al.}(2004)\citenamefont{Zacharia, Ulbricht,
  and Hertel}}]{ZachariaUH04}
\bibinfo{author}{\bibfnamefont{R.}~\bibnamefont{Zacharia}},
  \bibinfo{author}{\bibfnamefont{H.}~\bibnamefont{Ulbricht}}, \bibnamefont{and}
  \bibinfo{author}{\bibfnamefont{T.}~\bibnamefont{Hertel}},
  \bibinfo{journal}{Physical Review B} \textbf{\bibinfo{volume}{69}}
  (\bibinfo{year}{2004}).

\bibitem[{\citenamefont{Contreras-Camacho
  et~al.}(2004)\citenamefont{Contreras-Camacho, Ungerer, Boutin, and
  Mackie}}]{Contreras-CamachoUBM04}
\bibinfo{author}{\bibfnamefont{R.~O.} \bibnamefont{Contreras-Camacho}},
  \bibinfo{author}{\bibfnamefont{P.}~\bibnamefont{Ungerer}},
  \bibinfo{author}{\bibfnamefont{A.}~\bibnamefont{Boutin}}, \bibnamefont{and}
  \bibinfo{author}{\bibfnamefont{A.~D.} \bibnamefont{Mackie}},
  \bibinfo{journal}{Journal Of Physical Chemistry B}
  \textbf{\bibinfo{volume}{108}}, \bibinfo{pages}{14109}
  (\bibinfo{year}{2004}).

\bibitem[{\citenamefont{Watson et~al.}(2001)\citenamefont{Watson,
  Fechtenkotter, and Mullen}}]{WatsonFM01}
\bibinfo{author}{\bibfnamefont{M.~D.} \bibnamefont{Watson}},
  \bibinfo{author}{\bibfnamefont{A.}~\bibnamefont{Fechtenkotter}},
  \bibnamefont{and} \bibinfo{author}{\bibfnamefont{K.}~\bibnamefont{Mullen}},
  \bibinfo{journal}{Chem. Rev.} \textbf{\bibinfo{volume}{101}},
  \bibinfo{pages}{1267} (\bibinfo{year}{2001}).

\bibitem[{\citenamefont{Wick et~al.}(2000)\citenamefont{Wick, Martin, and
  Siepmann}}]{WickMS00}
\bibinfo{author}{\bibfnamefont{C.~D.} \bibnamefont{Wick}},
  \bibinfo{author}{\bibfnamefont{M.~G.} \bibnamefont{Martin}},
  \bibnamefont{and} \bibinfo{author}{\bibfnamefont{J.~I.}
  \bibnamefont{Siepmann}}, \bibinfo{journal}{Journal Of Physical Chemistry B}
  \textbf{\bibinfo{volume}{104}}, \bibinfo{pages}{8008} (\bibinfo{year}{2000}).

\bibitem[{\citenamefont{Errington and Panagiotopoulos}(1999)}]{ErringtonP99}
\bibinfo{author}{\bibfnamefont{J.~R.} \bibnamefont{Errington}}
  \bibnamefont{and} \bibinfo{author}{\bibfnamefont{A.~Z.}
  \bibnamefont{Panagiotopoulos}}, \bibinfo{journal}{Journal Of Physical
  Chemistry B} \textbf{\bibinfo{volume}{103}}, \bibinfo{pages}{6314}
  (\bibinfo{year}{1999}).

\bibitem[{\citenamefont{Linse}(1984)}]{Linse84}
\bibinfo{author}{\bibfnamefont{P.}~\bibnamefont{Linse}},
  \bibinfo{journal}{Journal Of The American Chemical Society}
  \textbf{\bibinfo{volume}{106}}, \bibinfo{pages}{5425} (\bibinfo{year}{1984}).

\bibitem[{\citenamefont{Linse et~al.}(1984)\citenamefont{Linse, Karlstrom, and
  Jonsson}}]{LinseKJ84}
\bibinfo{author}{\bibfnamefont{P.}~\bibnamefont{Linse}},
  \bibinfo{author}{\bibfnamefont{G.}~\bibnamefont{Karlstrom}},
  \bibnamefont{and} \bibinfo{author}{\bibfnamefont{B.}~\bibnamefont{Jonsson}},
  \bibinfo{journal}{Journal Of The American Chemical Society}
  \textbf{\bibinfo{volume}{106}}, \bibinfo{pages}{4096} (\bibinfo{year}{1984}).

\bibitem[{\citenamefont{Claessens et~al.}(1983)\citenamefont{Claessens,
  Ferrario, and Ryckaert}}]{ClaessensFR83}
\bibinfo{author}{\bibfnamefont{M.}~\bibnamefont{Claessens}},
  \bibinfo{author}{\bibfnamefont{M.}~\bibnamefont{Ferrario}}, \bibnamefont{and}
  \bibinfo{author}{\bibfnamefont{J.~P.} \bibnamefont{Ryckaert}},
  \bibinfo{journal}{Molecular Physics} \textbf{\bibinfo{volume}{50}},
  \bibinfo{pages}{217} (\bibinfo{year}{1983}).

\bibitem[{\citenamefont{Evans and Watts}(1976)}]{EvansW76}
\bibinfo{author}{\bibfnamefont{D.~J.} \bibnamefont{Evans}} \bibnamefont{and}
  \bibinfo{author}{\bibfnamefont{R.~O.} \bibnamefont{Watts}},
  \bibinfo{journal}{Molecular Physics} \textbf{\bibinfo{volume}{32}},
  \bibinfo{pages}{93} (\bibinfo{year}{1976}).

\end{thebibliography}

\end{document}